\documentstyle[emulateapj,onecolfloat]{article}

\input epsf
\font\tfa=cmr10 at 8.00pt
\lefthead{Metzler {\em et. al.}}
\righthead{Weak Gravitational Lensing and Cluster Mass Estimates}
 
\begin{document}
\twocolumn[
\title {Weak Gravitational Lensing and Cluster Mass Estimates}
\author {Christopher A. Metzler$^{1}$, Martin White$^{1,2}$,
Michael Norman$^{2,3}$ and Chris Loken$^{4}$}
\affil{$^{1}$Loomis Laboratory of Physics, University of Illinois,
1110 W. Green Street, Champaign, IL 61801--3080 USA}
\affil{$^{2}$Department of Astronomy, University of Illinois, 1002 W.
Green Street, Champaign, IL 61801}
\affil{$^{3}$Laboratory for Computational Astrophysics, National Center
for Supercomputing Applications, University of Illinois, 405 N. Mathews
Avenue, Urbana, IL 61801}
\affil{$^{4}$Department of Physics and Astronomy, University of Missouri,
Columbia, MO 65211}

\begin{abstract}
\noindent
Hierarchical theories of structure formation predict that clusters of galaxies
should be embedded in a web like structure, with filaments emanating from them
to large distances.  The amount of mass contained within such filaments near a
cluster can be comparable to the collapsed mass of the cluster itself.
Diffuse infalling material also contains a large amount of mass.
Both these components can contribute to the cluster weak lensing signal.
This ``projection bias'' is maximized if a filament lies close to the
line-of-sight to a cluster.
Using large--scale numerical simulations of structure formation in a
$\Lambda$--dominated cold dark matter model, we show that
the projected mass typically exceeds the actual mass by several tens of
percent.
This effect is significant for attempts to estimate cluster masses through
weak lensing observations, and will affect weak lensing surveys aimed at
constructing the cluster mass function.
\end{abstract}

]

\rightskip=0pt
\section{Introduction}

Clusters of galaxies are excellent cosmological probes.  Their size suggests
that their constituents provide a fair sample of the universe.
Their structure and hydrodynamic state provide information on their formation
and evolution, and thus upon models of structure formation.
Measurements of the abundance of clusters of a given mass allows constraint of
the amplitude of mass fluctuations in the universe; measurements of abundance
evolution can be used to constrain the mass density $\Omega_{\rm m}$.

Many of these approaches depend upon some knowledge of the mass, or mass
distribution, of the cluster.  Most techniques for measuring cluster
masses are based upon some equilibrium
assumption which relates the cluster mass to
an observable such as the temperature of the intracluster plasma
or the velocity dispersion of cluster galaxies.
Recently, however, it has become feasible to measure the surface density
distribution of a cluster through observations of weak gravitational lensing
of the background galaxy field by the cluster.
An attractive quality of this technique is that no assumptions about the
dynamical or thermodynamical state of the cluster components need be made.
In other words, weak lensing analyses probe the mass distribution directly.

However, analyses of the mass distribution of a cluster drawn from
weak lensing observations are not without problems (for a recent review,
see Mellier~\cite{Mellier}).
Many of these relate to details of the procedure adopted to go from the
observed ellipticity distribution to the mass, or from instrumental effects.
We will not discuss these in this {\em Letter}.
We are interested here in the degree to which weak lensing mass estimates
of clusters are affected by lensing from material outside but nearby the
cluster.  This is a source of systematic error which is not well understood
(though it has been alluded to in earlier work,
e.g.~Miralda-Escude~\cite{ME}; Wambsganss, Cen \& Ostriker~\cite{WamCenOst}).

Since clusters form in overdense regions, the volume surrounding a cluster
is likely to contain infalling overdense matter (Gunn \& Gott~\cite{GunGot}).
This infalling matter could add to an observed lensing signal and result in
an overestimate of the cluster mass.
It is possible that this bias could be quite severe.  In modern hierarchical
models of structure formation, such as the Cold Dark Matter (CDM) model,
numerical simulations imply that clusters form primarily at the intersections
of filaments in a web of cosmic structure, accreting additional diffuse
mass and smaller collapsed objects that drain along these filaments.
It is thus reasonable to expect a beaded filamentary structure surrounding
most clusters of galaxies.
Tentative observational evidence of filamentary structure near clusters
has been reported recently
(Kull \& Boehringer~\cite{KulBoe}, Kaiser et al.~\cite{MS0302}).
A filament lying near the line-of-sight will also lens the background
galaxies, and therefore contribute spuriously to the lensing signal.
If the observed lensing signal were attributed solely to the cluster,
the inferred cluster mass could be much larger than its actual mass.

In this {\em Letter}, we use mock clusters from numerical simulations
to explore the significance of the systematic mass overestimation
induced by the additional lensing signal of both diffuse and
filamentary material near the cluster.  We find that this effect can
be quite significant and must be considered when evaluating the results
of lensing mass reconstruction techniques.  In the next section, we
describe the numerical data and our procedure for evaluating the
errors introduced into cluster mass estimates by nearby material.
Section 3 describes the results of our analyses.  We discuss these
results and outline plans for future study at the end.

\section{Method}

\begin{table}
\begin{center}
\begin{tabular}{lcccc}
& & \multicolumn{3}{c}{Number} \\
& & 0 & 2 & 4 \\
$R_{\rm sphere}$    & $(h^{-1}{\rm Mpc})$  & 12.9 & 15.1 & 14.9 \\
$r_{200}$           & $(h^{-1}{\rm Mpc})$  & 3.14 & 2.76 & 2.60 \\
$M_{200}$           & $(h^{-1} 10^{15} M_{\sun})$   & 2.16 & 1.47 & 1.23 \\
$M_{>70}$           & $(h^{-1} 10^{15} M_{\sun})$   & 2.41 & 3.11 & 2.35 \\
$M_{>10}$           & $(h^{-1} 10^{15} M_{\sun})$   & 3.24 & 4.24 & 3.37 \\
$M_{\rm tot}$       & $(h^{-1} 10^{15} M_{\sun})$   & 3.89 & 5.09 & 4.15 \\
\end{tabular}
\end{center}
\caption{\tfa
Parameters for the 3 simulated clusters discussed in this
{\em Letter}.  $R_{\rm sphere}$ is the size of the sphere, centered on
the cluster, which we consider in this work, $r_{200}$ is the 3D radius
defined in Eq.~\protect\ref{eq:r200def} and $M_{200}$ the mass enclosed.
$M_{>70}$ and $M_{>10}$ are masses cutting out particles above thresholds
of 70 and 10 times local density respectively (see text).
$M_{\rm tot}$ is the total mass in the sphere.}
\label{tab:numbers}
\end{table}

Weak lensing mass reconstruction techniques produce a map of shear or
convergence.  These are integrals of the mass along the line-of-sight
times a projection kernel.
This kernel is quite wide in the redshift direction, scaling as
$D_LD_{LS}/D_S$ where $D_L$ is the distance from the observer to the lens,
$D_S$ from the observer to the source and $D_{LS}$ from the lens to the
source\footnote{For a distribution of source distances, one takes an
appropriate integral of this expression.} (Mellier~\cite{Mellier}).
Under the assumption that the cluster is the most massive object along the
line-of-sight and is well localized in space (the thin-lens approximation),
the convergence map is proportional to the projected surface density map
of the lensing cluster itself.

Any additional mass located near the cluster and along the line-of-sight
will also contribute to the lensing signal.  Since the kernel is such
a slowly varying function of distance, material even large distances from
the cluster will contribute within the thin lens approximation.
For a source at $z=1$ and a cluster at $z\sim0.5$ the kernel changes by
only 1\% within $\pm40\,h^{-1}\,$Mpc of the cluster in a universe with
$\Omega_{\rm m}=0.3=1-\Omega_\Lambda$, with similar results in other
cosmologies.
As a result, weak lensing observations will probe the projected density of
a cluster {\em plus\/} all of the material in its vicinity.
Note that this ``nearby'' material is essentially ``at'' the redshift of
the cluster for the purposes of lensing, and so cannot be distinguished
by using extra information such as source redshifts.

To study the effect of the surrounding mass upon the projected mass
inferred from lensing observations of the simulated clusters, we have examined
the mass distribution around several clusters of galaxies extracted from a
large cosmological simulation.
The simulated clusters
were taken from the X-Ray Cluster Data Archive of the
Laboratory for Computational Astrophysics of the National Center for
Supercomputing Applications (NCSA), and the Missouri Astrophysics Research
Group of the University of Missouri.
To produce these clusters, a particle-mesh N-body simulation incorporating
adaptive mesh refinement was performed
in a volume $256\,h^{-1}$Mpc on a side.
Regions where clusters formed were identified; for each cluster, the
simulation was then re-run (including a baryonic fluid) with finer
resolution grids centered upon the cluster of interest.
In the adaptive mesh refinement technique, the
mesh resolution dynamically improves as needed in high-density regions.
The ``final'' mesh scale at the highest resolution was $15.6\,h^{-1}$ kpc,
allowing good
resolution of the filamentary structure around the cluster.  Inside
the cluster, the characteristic separation between the smallest--mass
particles, given by $d\,=\,\left(4\pi\ r_{200}^3/3N\right)^{1/3}$ with $N$
the number of particles inside the region, was approximately 86 kpc
for all three clusters examined here.
The code is described in detail in Norman \& Bryan (\cite{TheCode}).

The clusters used here were extracted at $z=0$ from simulations of a
$\Lambda$CDM model, with parameters
$\Omega_{\rm m}=0.3$, $\Omega_{\rm B}=0.026$, $\Omega_{\Lambda}=0.7$,
$h=0.7$, and $\sigma_8=0.928$.
In this {\em Letter\/} we observe these clusters as if they were at $z=0.5$.
In future work we plan to investigate the dependence of these results on
cosmology and on cluster redshift.

Since the number density of rich clusters is approximately
$\phi_{*}\sim 10^{-5}$ Mpc$^{-3}$, the typical separation between them
is $\phi_{*}^{-1/3}\sim 40\,h^{-1}$Mpc.
This is a characteristic scale for filaments: volumes containing a cluster
and with one-dimensional extent $\sim 40\,h^{-1}$Mpc should also contain much
of the nearby filamentary structure.
Three such volumes, containing a rich cluster (Clusters 0, 2 and 4) as
well as satellites and
filaments, were extracted from the archive.  We selected clusters that did
not appear to be mergers or have a large secondary mass concentration nearby.
Such systems might be excluded observationally from studying, for instance,
the galaxy line-of-sight velocity distribution.
For each volume, the ``center'' of our cluster was determined using a
maximum-density algorithm.
As the extracted volumes were not spherical, it was possible that some
lines of sight could contain more mass than others simply by geometry.
To avoid such biases we restrict our analysis to particles that lay within the
largest sphere, centered on the cluster, which was contained entirely within
the extracted volume.
The radii of these spheres, $R_{\rm sphere}$, are listed in
Table~\ref{tab:numbers}, along with other properties of the clusters.
Note that these radii are large compared to the projected values of $r_{200}$
obtained for each cluster; thus no significant radial surface density gradient
is introduced by a decreasing chord length through the sphere with radius.
It is also important to note that since our volumes are by necessity
limited, our results should be interpreted as a {\em lower limit\/} to the
size of the effect; the magnitude of the lensing kernel is still significant
at the edge of our spherical volume.

\begin{figure}[t]
\centerline{\epsfxsize=3.3in \epsfbox{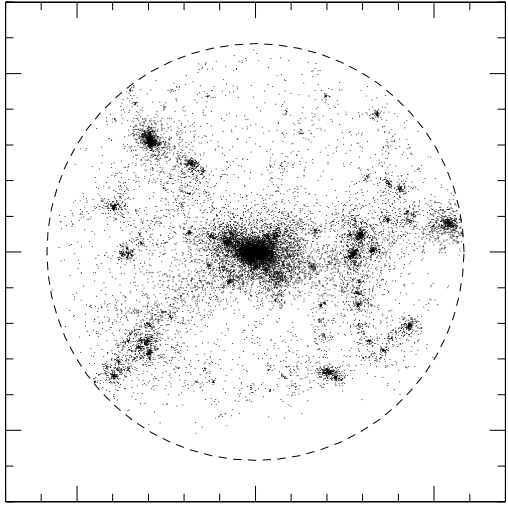}}
\caption{\tfa
Cluster 4, with a fraction of the particles projected onto
the $x-z$ plane.  Note the filamentary structure, with clumps beading
up in the filaments.  The dashed circle marks the sphere of radius
$R_{\rm sphere}$ to within which we have restricted our analysis
(see text).}
\label{fig:cl004}
\end{figure}

Each of the clusters we examined was surrounded by a large amount of mass.
Most of this material appeared by eye to be collapsed into ``beads'' along a
filamentary structure, although a small number of clumps could be found
outside the filaments.
We show a projection of a fraction of the points from the simulation
of cluster 4 in Fig.~\ref{fig:cl004}.
The filamentary structure and satellites are easily evident.  Note that
this filamentary structure extends
well beyond our radius $R_{\rm sphere}$.
No single projection can show the full 3D nature of the structure,
in which the filamentarity is even more apparent.
Since much of this mass is at low density it is unlikely it would be a site
for galaxy formation or otherwise emit light.  Thus this structure would not
be easy to constrain by observations of redshifts near the cluster.

We observed each of the three selected clusters from $10,000$ randomly
chosen viewing angles.
For each cluster and viewing angle, the projected surface density map was
constructed and used to estimate $r_{200}$, the radius within which the mean
interior density contrast is 200.
In three dimensions, this radius is defined in terms of the enclosed mass by
\begin{equation}
  M\left(<r_{200}\right)=200\times \left({4\pi\over3}\right)
    \Omega_{\rm m}\, \rho_{\rm crit}\, r_{200}^3.
\label{eq:r200def}
\end{equation}
The projected estimate of $r_{200}$ was extracted from the surface
density map by considering the radius of the circle, centered on the
cluster, which contained the amount of mass given by
Eq.~(\ref{eq:r200def}) above, i.e.
\begin{equation}
 \int_0^{2\pi}{\rm d}\theta \int_0^{r_{200}} R\, {\rm d}R
\,\ \Sigma\left(R,\theta\right)
   = M\left(<r_{200}\right)
\end{equation}
with $\Sigma\left(R,\theta\right)$ the surface density on the map in terms of a
two-dimensional radius $R$.  This radius was compared to the cluster's
true $r_{200}$, extracted from the three--dimensional mass distribution.
The ratio of the projected mass to true mass is given simply by the cube
of the ratio of the estimated value of $r_{200}$ to the true value.
For each cluster, a value of this ratio was obtained for each viewing angle.

\section{Results}

\begin{figure}[t]
\centerline{\epsfysize=9cm \epsfbox{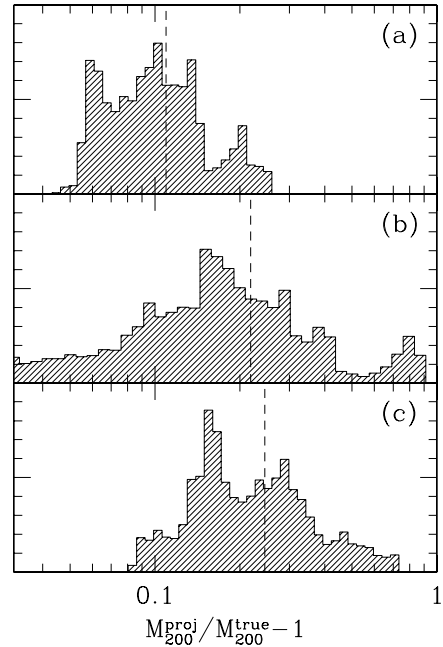}}
\caption{\tfa
The ratio of projected to actual mass within $r_{200}$ for our
3 simulated clusters.  In all cases nearby mass (primarily in filaments)
has biased the projected mass distribution to higher values.  The typical
bias is a few tens of percent (b-c), with less of an effect on massive
clusters (a).  The vertical dashed line marks the mean of the distribution.}
\label{fig:mass}
\end{figure}

Our main result is displayed in Fig.~\ref{fig:mass}, where we show the
distribution of projected vs.~``true'' cluster mass in each of the three
simulated clusters.
We have checked that the features in the histogram do not come from shot noise
due to discrete particles in the simulation.  However, the ``spikiness''
{\em is\/} due to a discrete number of objects in the neighborhood of the
cluster.  A small lump of matter near the cluster will project entirely within
$r_{200}$ for a fraction of the lines of sight.
For any such line of sight, the effect on the projected value of $r_{200}$
is identical.

We expect the ratio $M_{200}/M_{\rm true}$ to be greater than unity since
only additional mass can be included in the projection.  The size of the
smallest offset from unity for clusters 0 and 4 is approximately twice what
would be expected for material uniformly distributed at the mean density.
This suggests that matter near the cluster is itself clustered and at higher
than mean density.
The width of the histogram in Fig.~\ref{fig:mass}, as a fraction of the true
mass, depends on the true mass of the cluster.
Though we have only a few clusters in this study, it appears that the mass in
nearby material is not proportional to the cluster mass, thus the relative
effect of this material is smaller the larger the cluster.
The total mass in the sphere, $M_{\rm tot}$, is also listed in
Table~\ref{tab:numbers} for reference.

The signal shown in Fig.~\ref{fig:mass} comes from (primarily filamentary)
material outside the cluster and is {\em not\/} the well known projection
effect arising from cluster asphericity.
To verify this, we repeated the procedure described above for a subset of
particles aimed at selecting the cluster alone.  This was done by first
selecting out particles with a local density contrast of greater than 70
(chosen because density profiles near $r^{-2}$ reach a local density contrast
near 70 at a mean interior density contrast of 200); a small sphere
containing the cluster but little nearby material was then cut out
of this subset.  The histogram produced by viewing the clearly prolate
cluster at a large number of randomly chosen viewing angles produced
a much narrower distribution, with a maximum offset of less than $10\%$
in the mass ratio and a mean offset of approximately half that value.

\begin{figure}[t]
\centerline{\epsfxsize=3.3in \epsfbox{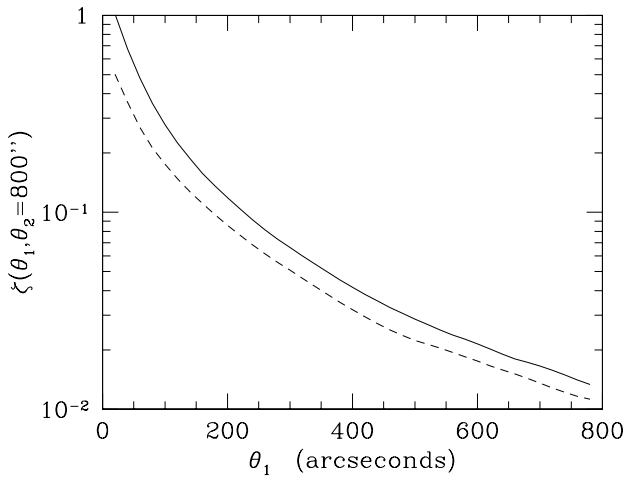}}
\caption{\tfa
The profile $\zeta(r_1,r_2)$ (see text) for cluster 4 along two
lines of sight with extremal values of $r_{200}$.
We have used $r_2=800''$ in making this figure.  Such a large radius is
not currently achievable observationally, but it minimizes the impact of
objects nearby the cluster and provides a lower limit on the effect.
Reducing this radius by a factor of two does not change the result.
Note that both profiles appear smooth and well behaved, even though they
differ in mass by a factor of 1.7.}
\label{fig:zeta}
\end{figure}

While it is beyond the scope of this {\em Letter\/} to perform a detailed
modelling of any observational weak lensing strategy, we show in
Fig.~\ref{fig:zeta} two sample profiles obtained from aperture densitometry
on our noiseless projected mass maps.
Specifically, for Cluster 4, we show the profile along the lines of sight
giving the largest and smallest $r_{200}$, for comparison.
The $\zeta$ statistic is the mean value of the convergence, $\kappa$, within
a disk of radius $r_1$ minus the mean value within an annulus
$r_1\le r\le r_2$ (Fahlman et al.~\cite{Fahlman}, Kaiser~\cite{K95}).
Here we calculate $\zeta$ directly from the projected mass, though
observationally it would be computed from the tangential shear.
We have taken $r_2=800''$.
Such a large radius is not (currently) achievable observationally, but it
minimizes the impact of objects nearby the cluster and provides a lower
limit on the size of the projection effect.
We have explicitly checked that reducing the radius to half this value does
not change our result.

In calculating the convergence $\kappa$, the cluster was again assumed
to be at a redshift of $0.5$, with the lensed sources at a redshift of
$1.0$.  In any real observation, of course, the lensed sources will
span a range of redshifts.  For material very close to the cluster, such
as here, this will not affect our conclusions, and the error introduced by
incorrectly modelling the redshift distribution of the background sources
is not the subject of this work.
In addition to being easy to estimate, the $\zeta$ statistic is robust and
minimizes contamination by foreground mass (Mellier~\cite{Mellier}) because
it is insensitive to the sheet mass degeneracy.  This does not, however,
remove the sensitivity to {\em clustered\/} material, as can be seen in
Fig.~\ref{fig:zeta}.
The mass which would be inferred from Fig.~\ref{fig:zeta} along two lines
of sight differ by a factor of $1.7$.

While the distribution of the projection effect varies from cluster
to cluster, it seems clear that positive biases in projected mass of
$20\%$ are typical, with biases above $30\%$ not uncommon.  Furthermore,
we emphasize again that these estimates are in fact {\em lower limits\/};
some lines of sight through the untruncated volume produced overestimates
as large as $80\%$ or more.  While appropriate modeling of a mean density
profile outside $r_{200}$ (drawn perhaps from simulations such as these)
can be used to produce an unbiased estimator, the width of these histograms
implies a large amount of scatter around such an estimator of the true
(unprojected) mass.  It is clear that this effect can be quite significant and
must be taken into account when attempting to understand the results of mass
reconstruction analyses.

\section{Conclusions}

Clusters of galaxies are part of the large-scale structure of the universe
and observations of them should be considered within this context.
The filamentary structure near a cluster can
contain a reasonable fraction of the mass
of a cluster in tenuous material.  Should a filament lie close to the
line-of-sight to a cluster it will contribute to the weak lensing signal and
positively bias the projected mass.
We have shown that such a bias could easily be 30\% (see Fig.~\ref{fig:mass}).

Weak lensing remains one of the best methods for determining the mass of
clusters of galaxies.  However methods which obtain the mass from estimates
of the projected surface density must consider the effect outlined in this
{\em Letter}.  This is clearly of particular significance for attempts to
determine the mass function of clusters directly through surveys of weak
lensing-determined masses.

We have not attempted to address the detailed question of how much this
filamentary signal would affect a {\em particular\/} reconstruction
algorithm; the answer is no doubt algorithm dependent.
We hope to return to this issue in future work, as well
as to consider the effect of cosmological model and evolution with cluster
redshift.

The authors would like to thank Greg Bryan and Greg Daues for assistance
in understanding the archive data, and Gordon Squires and Albert Stebbins
for useful conversations on lensing.
This research was supported by the NSF.

\end{document}